\documentclass[arxiv]{melba}

\usepackage{amsmath,amsfonts}
\usepackage{multirow}
\usepackage{graphicx}
\usepackage{adjustbox}
\usepackage{array}
\usepackage{subcaption}

\doi{https://doi.org/10.59275/j.melba.2024-AAAA}
\melbaauthors{Omar Ali H, Abraham R, Desoubeaux G, Fahal A, Tauber C}  
\melbayear{2024}  
\datesubmitted{m1/yyyy}  
\datepublished{m2/yyyy}  

\ShortHeadings{MyData: A Comprehensive Database of Mycetoma Tissue Microscopic Images for Histopathological Analysis}{Omar Ali H, Abraham R, Desoubeaux G, Fahal A, Tauber C}

\title{MyData: A Comprehensive Database of Mycetoma Tissue Microscopic Images for Histopathological Analysis}

\author{
	\firstname Hyam \surname Omar Ali\aff{1,2,3,4},
	\name Romain Abraham\aff{4},
	\name Guillaume Desoubeaux\aff{5,6},
	\name Ahmed Fahal\aff{2},
	\name Clovis Tauber\aff{3},
}
\affiliations{
	\num 1 \addr Faculty of Mathematical Sciences, University of Khartoum, 11111, Khartoum, Sudan \\
	\num 2 \addr The Mycetoma Research Centre, University of Khartoum, 11111, Khartoum, Sudan \\
	\num 3 \addr U1253 iBrain, University of Tours, Inserm, 37032, Tours, France \\
	\num 4 \addr CNRS U7013 Institute Denis Poisson, University of Orleans, 45067, Orleans, France \\
	\num 5 \addr Parasitology and Mycology Department, Bretonneau Hospital, 37032, Tours, France \\
	\num 6 \addr INSERM U1100, University of Tours, 37032, Tours, France
}

\abstract{
Mycetoma is a chronic and neglected inflammatory disease prevalent in tropical and subtropical regions. It can lead to severe disability and social stigma. The disease is classified into two types based on the causative microorganisms: eumycetoma (fungal) and actinomycetoma (bacterial). Effective treatment strategies depend on accurately identifying the causative agents. Current identification methods include molecular, cytological, and histopathological techniques, as well as grain culturing. Among these, histopathological techniques are considered optimal for use in endemic areas, but they require expert pathologists for accurate identification, which can be challenging in rural areas lacking such expertise. The advent of digital pathology and automated image analysis algorithms offers a potential solution. This report introduces a novel dataset designed for the automated detection and classification of mycetoma using histopathological images. It includes the first database of microscopic images of mycetoma tissue, detailing the entire pipeline from species distribution and patient sampling to acquisition protocols through histological procedures. The dataset consists of images from 142 patients, totalling 864 images, each annotated with binary masks indicating the presence of grains, facilitating both detection and segmentation tasks.}


\begin{document}

\twocolumn[\maketitle]

\section{Background}\label{introduction}
	\enluminure{M}{ycetoma} is a significantly neglected tropical disease characterised by chronic, disabling subcutaneous granulomatous inflammation. It is caused by a variety of microorganisms, leading to its classification into two types: eumycetoma, caused by fungi, and actinomycetoma, caused by bacteria \citep{who_mycetoma}. The most common species responsible for eumycetoma is Madurella mycetomatis (M. mycetomatis). In contrast, common species for actinomycetoma include Streptomyces somaliensis (S. somaliensis), Actinomadura madurae (A. madurae), Actinomadura pelletieri (A. pelletieri), and Nocardia brasiliensis (N. brasiliensis) \citep{van2013global,van2014mycetoma, zijlstra2016mycetoma,hay2021mycetoma}.
		
	\noindent Mycetoma occurs worldwide, but the overall global burden of the disease is still unclear. The infection is predominantly reported from tropical and subtropical regions, with most cases occurring in the so-called "Mycetoma belt" \citep{who_mycetoma}. The highest prevalence has been reported in Sudan, India, and Mexico. Sudan is considered the mycetoma epicenter,  with the Mycetoma Research Centre (MRC) at the University of Khartoum, a WHO Collaborating Centre on Mycetoma, reporting 355 new cases annually \citep{fahal2015mycetoma}. However, this number likely underestimates the actual number of cases, as it only includes patients who were able to access medical care at the MRC.
	
	\noindent Young adults, especially males aged 20 to 40 in remote rural areas, are the most vulnerable to mycetoma infection \citep{who_mycetoma, fahal2015mycetoma}. Women are less likely to be infected than men, with a ratio of 1:3. \citep{fahal2004mycetoma, fahal2015mycetoma}. The disease predominantly affects field labourers, agriculturalists, and herdsmen \citep{fahal2004mycetoma}. The lower extremities and hands are the most commonly infected areas \citep{van2013global, fahal2004mycetoma}. 
	
	\noindent In endemic areas, the initial diagnosis of mycetoma is often made clinically through physical examinations, though this can be challenging and inaccurate \citep{van2014mycetoma,emmanuel2018mycetoma}. To determine the appropriate treatment, further investigations to identify the causative agents are essential. Accurate identification is critical for effective treatment and minimising complications \citep{hay2021mycetoma}, typically requiring the use of laboratory-based diagnostic tools \citep{ahmed2017mycetoma,emmanuel2018mycetoma}. Histopathological techniques are considered an efficient, cost-effective, and time-saving method for diagnosing mycetoma in these regions \citep{van2014merits}.
	
	\noindent Over the past decades, medical image analysis models, particularly those utilising deep learning, have become a crucial approach in the medical field. However, no model has been developed specifically for diagnosing mycetoma, likely due to the neglect of this condition in both health and computational research. A dedicated database is needed to facilitate the development of a mycetoma image analysis model. Therefore, we created \textbf{\textit{MyData}}, the first database of histopathological microscopic images of mycetoma, following specific and consistent standards. This database aims to serve as a valuable resource for developing AI-based diagnostic tools, enhancing diagnostic accuracy, and improving patient outcomes. MyData can be used by the scientific community to model and analyses mycetoma histological characteristics, providing insights to develop more effective diagnostic parameters and strategies.  
	
	\noindent In this communication, Section \ref{introduction} introduces the work, while Section \ref{sec_2} discusses the epidemiological distribution of mycetoma causative agents. Section \ref{sc_pop} details the study population, including sample selection and diagnosis. Section \ref{histo} outlines the histopathological process pipeline and the preparation of microscopic images, which is the main focus of this work. A comprehensive description of the proposed mycetoma database, along with a summary of the data and labels, is presented next. The final section concludes the paper.
	 
\section{The distribution of Mycetoma Causative Agents and Organisms}\label{sec_2}
\noindent To introduce the mycetoma database, it is crucial to outline the epidemiological distribution of mycetoma causative agents to ensure the database accurately represents mycetoma cases globally. 

\begin{figure}[h!]
	\centering
	\includegraphics[width=1.0\linewidth, height=.2\textheight]{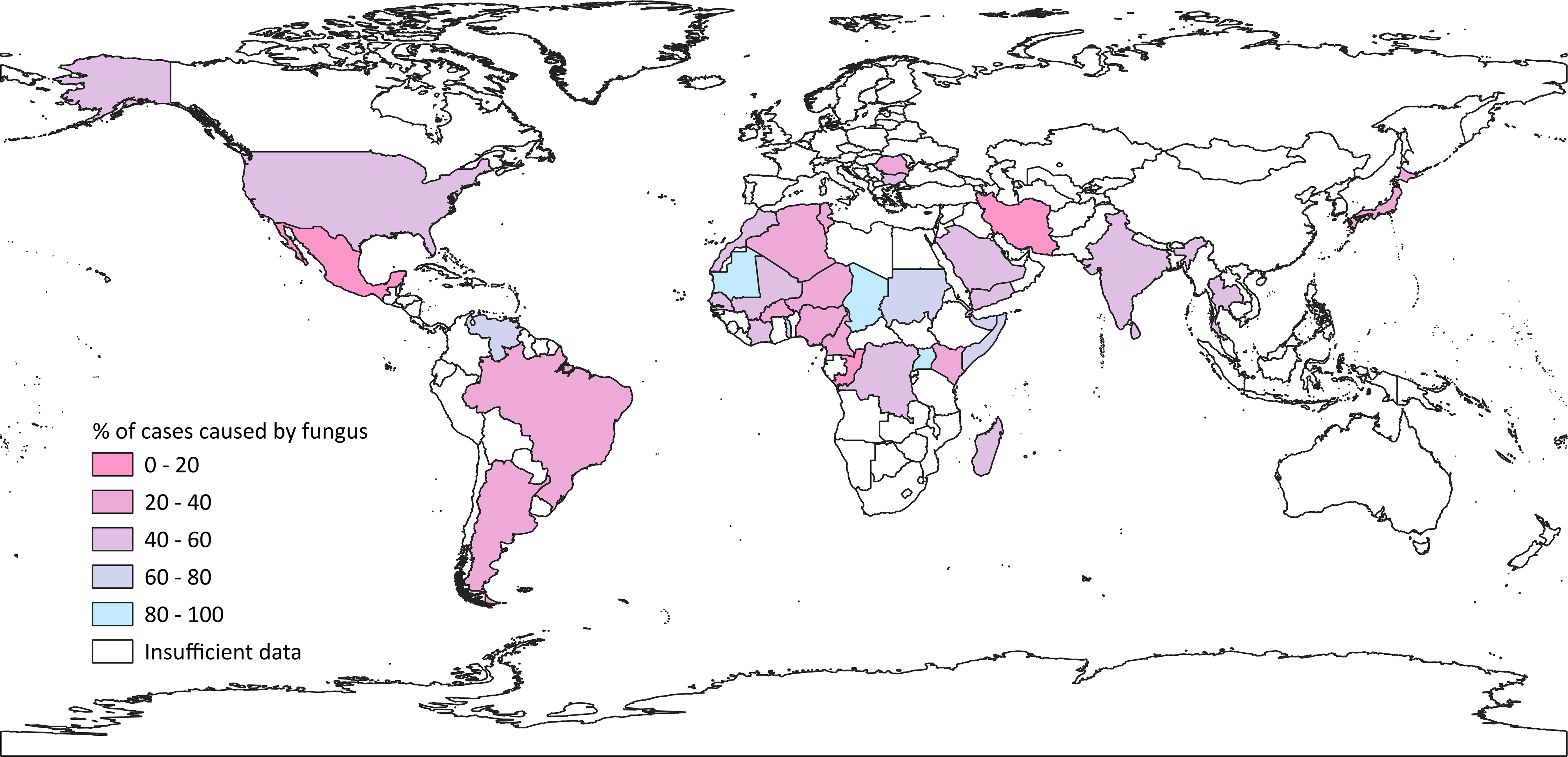}
	\caption{Prevalence of eumycetoma and actinomycetoma \citep{emery2020global}}
		\label{map}
\end{figure}

\noindent Approximately 40\% of mycetoma cases worldwide are caused by eumycetoma \citep{van2013global, dndi_mycofacts}. However, eumycetoma is more prevalent in certain regions, such as central and eastern African countries, while actinomycetoma is more common in the Americas and the Middle and Far East \citep{emery2020global}. Additionally, the distribution of eumycetoma and actinomycetoma can vary within the same region and between neighbouring countries, Figure \ref{map}. For instance, most countries report mixed cases of both types, but some countries have a higher proportion of one type. In Sudan, for example, $ 73\% $ of cases are eumycetoma, whereas in India, this percentage drops to $ 42\% $, and in Mexico, only $ 3\% $ of cases are eumycetoma \citep{emery2020global}.

\noindent While the pathogens causing mycetoma are either fungal or bacterial, overall mycetoma can be caused by up to 70 different causative organisms \citep{who_mycetoma}. Globally, the prevalence of these agents has been documented \citep{van2013global}, most recently \citep{emery2020global}. In this study, Madurella mycetomatis (MM) and Nocardia species were the predominant causative organisms, likely due to the study focused on qualitative rather than quantitative data. Given the uneven distribution of mycetoma types even within individual countries, the mapping performed in this study provides a notable overview of the distribution of mycetoma-causative organisms. In general, the most common causative organisms include Madurella mycetomatis (MM), Actinomadura madurae (AMM), Streptomyces somaliensis (SS), Actinomadura pelletieri (AMP), Nocardia brasiliensis, and Nocardia asteroides. However, Nocardia species are more prevalent in South America and Asia and less common in Europe and Africa \citep{van2013global}.

\noindent The MyData database sample was collected at the MRC, so the types of mycetoma represented reflect the geographical distribution in Sudan, where eumycetoma is more prevalent. The database includes organisms such as Madurella mycetomatis (MM), Actinomadura madurae (AMM), Streptomyces somaliensis (SS), and Actinomadura pelletieri (AMP). However, it does not include Nocardia species, as they are rarely observed in Sudan. While the inclusion of specific mycetoma organisms in our database does not impact treatment options, it can influence the prognosis of the disease. This information may be useful for purposes beyond developing AI models for identifying causative agents.

\section{MyData Population} \label{sc_pop}
\subsection{Sample collection and Selection criteria}
\noindent The initial MyData database comprised 180 patients with clinically confirmed mycetoma infection, who were either seen at the MRC or identified through field surveys in Sudan. Surgical biopsies were collected from patients with various mycetoma types, durations, and clinical presentations. Thirty-eight patients were excluded from the study because their biopsies lacked grains, which are essential for a conclusive diagnosis of mycetoma, even if clinical symptoms are present. As a result, the MyData database includes samples from 142 patients with confirmed mycetoma infection. These patients were randomly selected from cases seen over the past five years to ensure homogeneity and diagnostic accuracy, regardless of age, sex, or race. Samples were selected to balance the eumycetoma and actinomycetoma classes, but the genus and species were not considered for the selection criteria. A summary of the MyData database population is provided in Table \ref{popu}.

	\begin{table}[h] 
    \centering
    \caption{The demographic of the studied population.}
    \label{popu}
    \resizebox{\columnwidth}{!}{%
\begin{tabular}{|c|c|c|}
	\hline
	\textbf{Characteristic} & \textbf{Category} & \textbf{Number} \\
	\hline \hline
	\multirow{1}{*}{Age in years} & $ 10-70 $ &  \\
	\hline
	\multirow{2}{*}{Sex} & Male & 89 \\
	& Female & $ 53 $ \\
	\hline
	\multirow{2}{*}{Type} & Eumycetoma & $ 80 $ \\
	& Actinomycetoma & $ 62 $ \\
	\hline
	\multirow{3}{*}{Site of infection} & Hands & $ 40 $ \\
	& Feet & $ 68 $ \\
	& Others & $ 34 $ \\
	\hline
	\multirow{4}{*}{Duration in years} &$  <1 $ & $ 18 $ \\
	& $ >1-5 $ & $ 69 $ \\
	& $ >5-10 $ & $ 46 $ \\
	& $ >10 $ & $ 9 $ \\
	\hline
	\multirow{3}{*}{Lesion size} & Small ($< 5$ cm in diameter) & $ 35 $ \\
	& Moderate ($ 5–10 $ cm) & $ 59 $ \\
	& Massive ($> 10$ cm in diameter) & $ 48 $ \\
	\hline
\end{tabular}
}
\end{table}

\subsection{Sample diagnosis}
\noindent Specialised microbiologists at the MRC conducted grain culture techniques to identify the mycetoma causative agents. We reviewed all histological tissue blocks along with their corresponding culture diagnoses to check for any misdiagnosis or incorrect identification of the causative agents. Any samples with conflicting diagnoses were excluded from the database.

\noindent Next, we examined all data on the genera and species responsible for mycetoma. Since fungal species are often misidentified using culture techniques \citep{van2013global}, we identified actinomycetoma at the species level and eumycetoma at the genus level. For some eumycetoma samples, we used molecular techniques to achieve species-level identification.

\noindent All the actinomycetoma samples in our database were identified as either Actinomadura madurae (AMM), Streptomyces somaliensis (SS), or Actinomadura pelletieri (AMP). They were determined using both grain culture and histopathological examination. For our eumycetoma data, 32 patients were assessed using both grain culture and histopathology and were classified as Madurella species (Mspp), Aspergillus species (Aspp), or Fusarium species (Fspp). For an additional 48 patients, a molecular technique was used to determine whether they were Madurella mycetomatis positive (MM+) or not (MM-). 
A summary of this data is provided in Table \ref{dist}.
	\begin{table}[h!]
	\centering
	\caption{The distribution of study samples.}
	\label{dist} 
	\resizebox{\columnwidth}{!}{%
	\begin{tabular}{|c|c|c|}
		\hline
		\textbf{Causative Agent} & \textbf{Species} & \textbf{No. Patients} \\ \hline \hline
		\multirow{5}{*}{Eumycetoma} & Madurella spp. (Mspp) & 22 \\
		& MM+                  & 29 \\ 
		& MM-                  & 19 \\ 
		& Aspergillus spp. (Aspp) & 8 \\ 
		& Fusarium spp. (Fspp) & 2 \\ \hline
		\multirow{3}{*}{Actinomycetoma} & Actinomadura pelletieril (AMP) & 9 \\  
		& Actinomadura madurae (AMM)    & 9 \\ 
		& Streptomyces somaliensis (SS) & 44 \\ \hline
	\end{tabular}
}
\end{table}

\section{Histopathology and Slides Preparation}\label{histo}
\begin{figure}[h!]
	\centering
	\includegraphics[width=0.9\linewidth, height=.3\textheight]{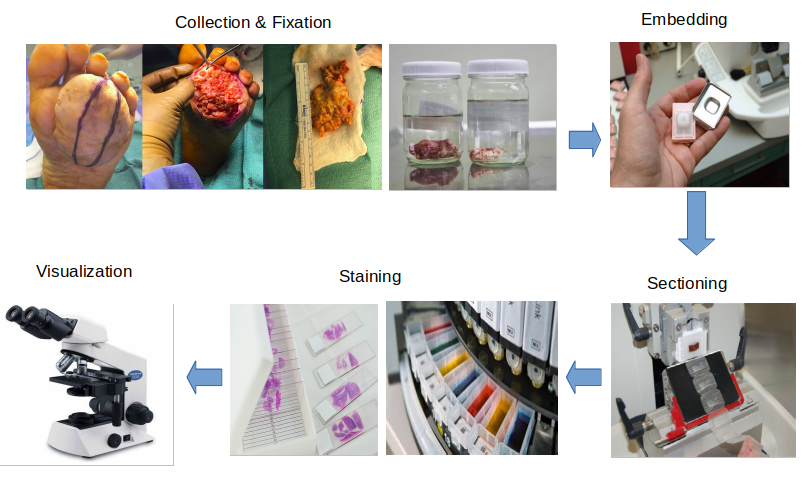}
	\caption{Comprehensive View of Histopathology Process}
	\label{hist_fig}
\end{figure}

This section outlines the essential steps involved in preparing tissue slides for histopathological examination, Figure \ref{hist_fig}. 

\noindent The process begins with collecting tissue samples from the site of investigation through surgery or biopsies, adhering to standard histological procedures. The preparation process involves three key steps \citep{suvarna2018bancroft}:

\begin{enumerate}
\item Tissue collection and Fixation:  Tissue was obtained via fine needle aspiration, tru-cut needle biopsy, or surgical biopsy. Larger biopsies provided more material for a comprehensive examination. The collected tissue was immersed in a fixative solution to prevent degradation and inhibit microorganism growth.

\item Embedding and Sectioning: The tissue was embedded in a supportive material, typically paraffin wax, to create solid blocks. These blocks were then sliced into thin sections, usually $(3-5)\mu$m thickness, using a microtome. The sections were placed in a heated water bath to melt the paraffin and smooth out wrinkles before being mounted on glass microscope slides for preservation and enhanced visual clarity.

\item Staining: To improve contrast and aid in identifying regions of interest (ROIs), the tissue sections were stained. Haematoxylin and Eosin (H\&E) stain, the most commonly used histological stain, was used. 
\end{enumerate}
\noindent These procedures were conducted at the microscopy facility of the University of Tours and according to standard protocols at Bretonneau Hospital in Tours, France. The prepared sections were then imaged using an optical light microscope to capture detailed microscopic images to document the grain morphology and the host tissue reaction and components.

\section{Preparation of Microscopic Images in MyData}\label{pre_mydat}
\subsection{Mycetoma Tissue Images Acquisition}
\begin{figure}[h!]
	\centering
	\includegraphics[width=0.4\linewidth, height=0.3\textheight]{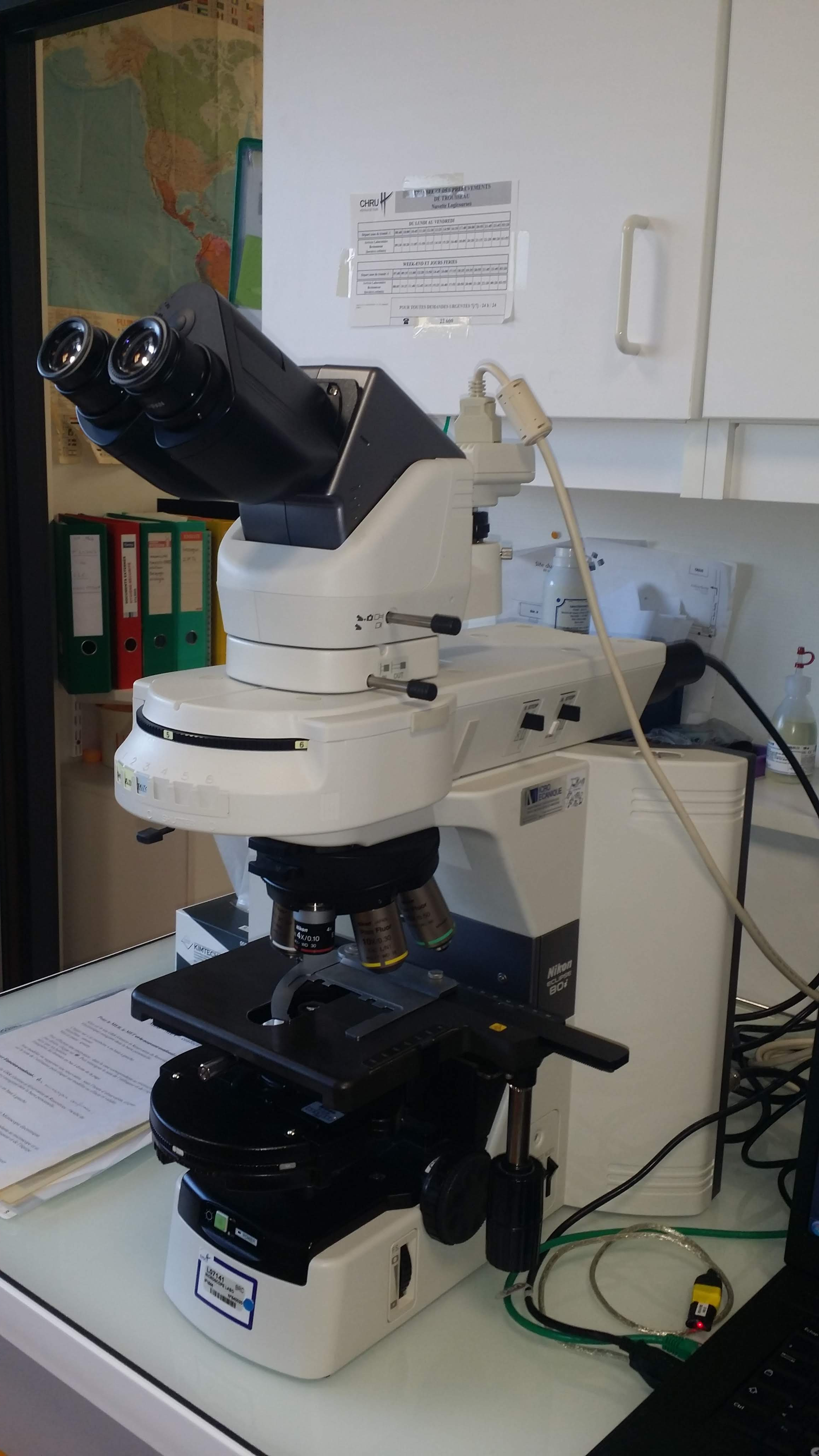}

	\caption{Digital Microscope}
	\label{micro}
\end{figure}

The tissue slides from the tissue blocks of 142 patients were handled with a unique reproducible acquisition protocol to ensure a uniform database. Microscopic images were captured in RGB color space using a Nikon Eclipse 80i digital microscope (Figure \ref{micro})  according to the conditions given in Table \ref{micro_table}. For each patient, a variety of grains were examined, with an average of six grains per patient. 

\begin{table}[h!]
	\centering
	\caption{The distribution of study samples.}
	\label{micro_table} 
	\begin{tabular}{|c|c|}
		\hline
		\textbf{Parameter} & \textbf{Value} \\ \hline \hline
		\multirow{3}{*}{Brightness control} & Knob 5/10 \\
		& ND8  On  \\
		& ND32  On  \\
		\hline
		Field diaphragm & Highest level \\ \hline
		Magnification  & $10\times $ \\ \hline
		Dimension and Quality  & $800\times 600$ \\ \hline
		Colour  & Enhance and white auto \\ \hline
		Field diaphragm knob  & Highest level \\ \hline
		Filter & 6 \\ \hline
		NCB11 Filter & Off \\ \hline
	\end{tabular}
\end{table}

\begin{figure*}[h!]
	\centering
	\begin{subfigure}[b]{0.45\textwidth}
		\centering
		\includegraphics[height=0.2\textheight]{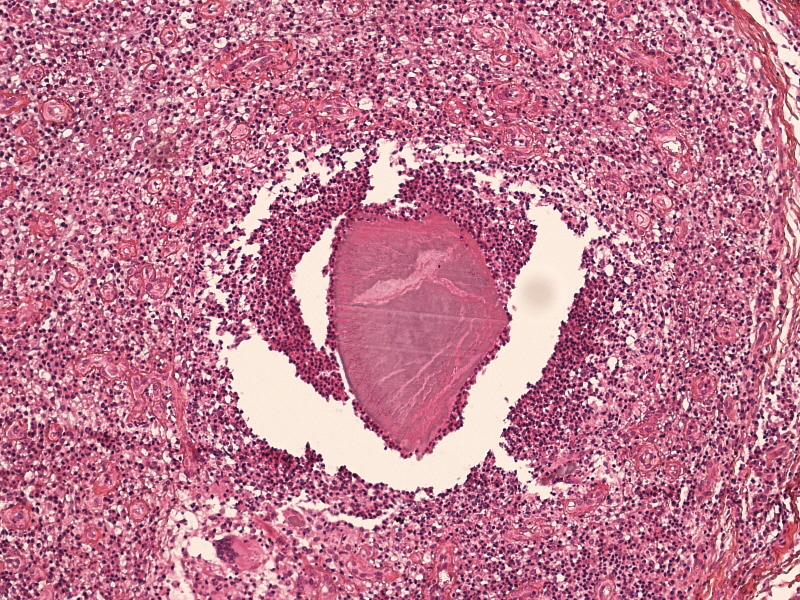}
		\caption{}
		\label{fig:one_grain}
	\end{subfigure}
	\begin{subfigure}[b]{0.45\textwidth}
		\centering
		\includegraphics[height=0.2\textheight]{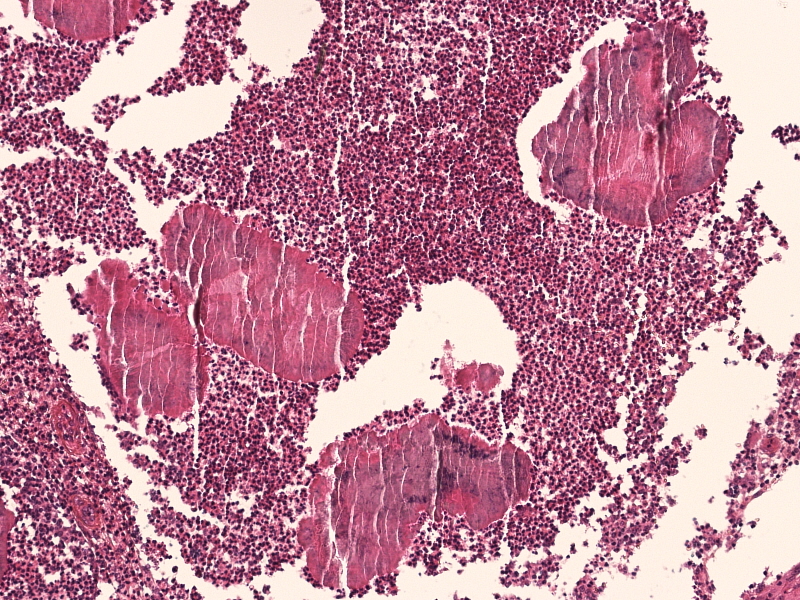}
		\caption{}
		\label{fig:many_grain}
	\end{subfigure}
	\caption{Mycetoma Microscopic Images. (a): Single grain. (b): Multiple grains.}
	\label{grains_no}
\end{figure*}

\noindent While the histopathological tissue slide screening aimed to capture one grain per field (Figure \ref{fig:one_grain}), some slides contained adjacent grains, leading to multiple grains appearing in a single image (Figure \ref{fig:many_grain}). In these instances, only one grain per image was included in the image.

\subsection{Ground-truth Segmentation of Mycetoma Grains}
\noindent For histopathological diagnosis, it is crucial to identify grains within the tissue to determine the causative agents and provide appropriate treatment. Consequently, mycetoma grains are considered the Regions of Interest (ROIs) in the tissue images. Manual segmentation of these grains was essential to create a comprehensive and accurate database of mycetoma tissue grains and their annotations. This database can then be utilised for computational tasks related to histopathological tissue diagnosis.

\noindent Grains from the histopathological  microscopic images of mycetoma-infected tissues were manually segmented using ImageJ software. This task was performed by an expert from the MRC with a strong background in image processing, ensuring high-quality and accurate annotations. Each image in the dataset was annotated by labelling pixels as either ROI or background, with pixels labelled as 1 representing the ROI and those labelled as 0 representing the background. These original images and their annotations are used for training and validating AI models.

\begin{figure*}[h!]
	\centering
\begin{subfigure}[b]{0.45\textwidth}
	\centering
	\includegraphics[height=0.2\textheight]{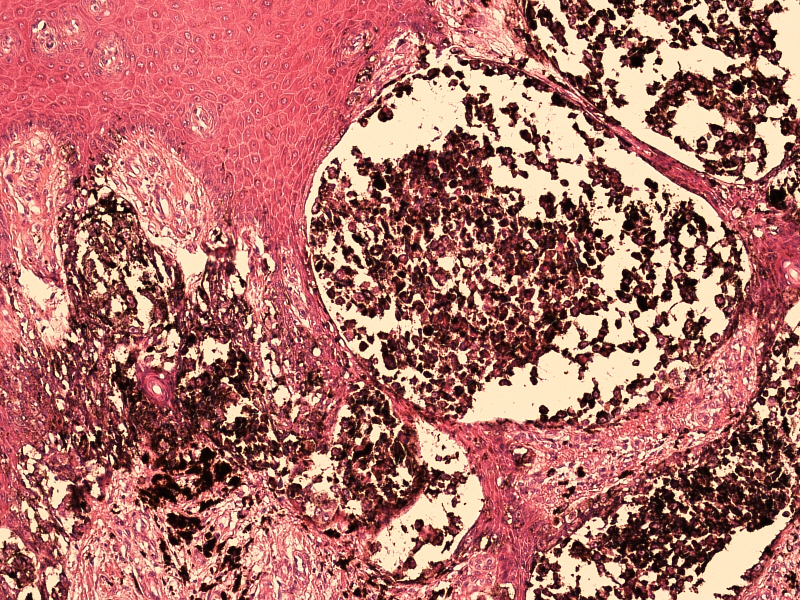}
	\includegraphics[height=0.2\textheight]{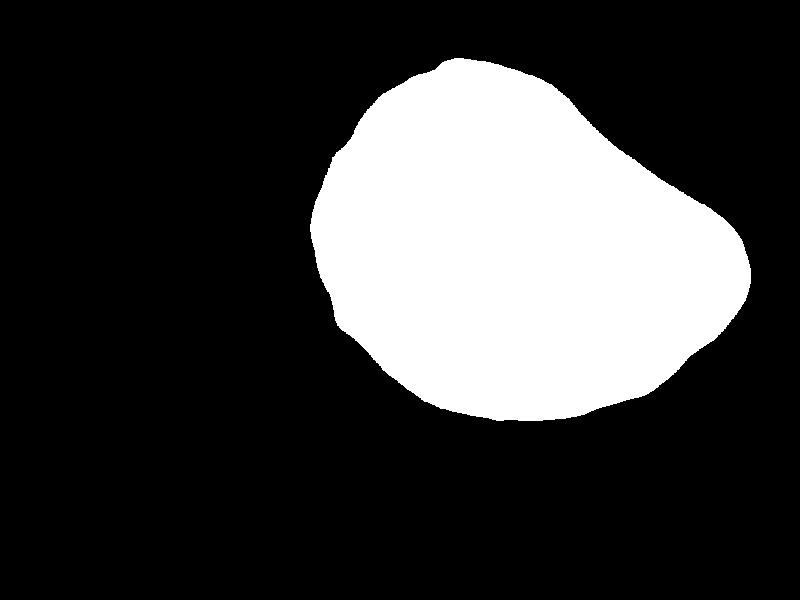}
	\caption{}
\end{subfigure}%
\hfill
\begin{subfigure}[b]{0.45\textwidth}
	\centering
	\includegraphics[height=0.2\textheight]{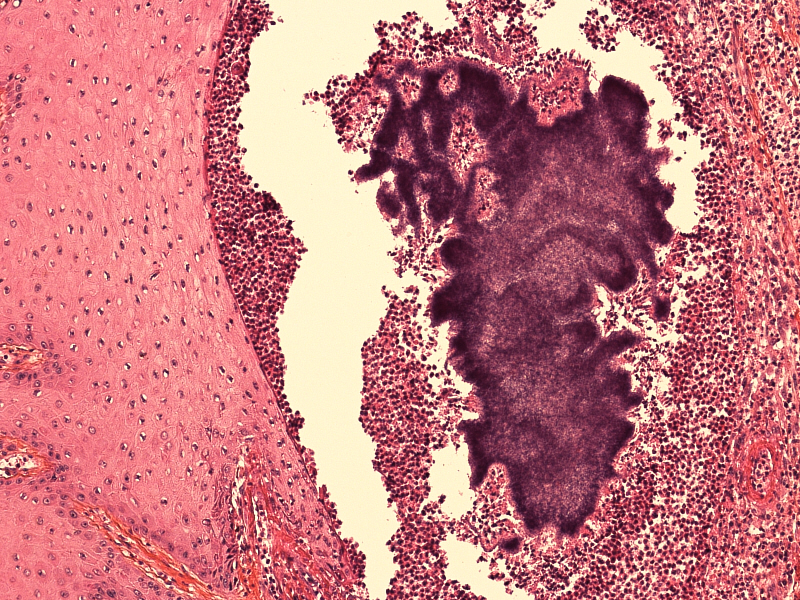}
	\includegraphics[height=0.2\textheight]{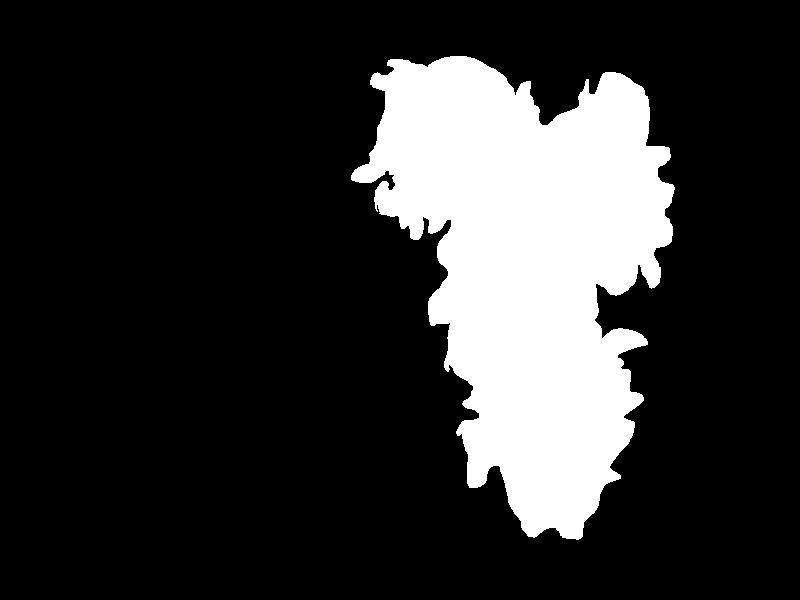}
	\caption{}
\end{subfigure}

\caption{Mycetoma Microscopic Images and its annotated grain. (a): Eumycetoma with fracture grain. (b): Actinomycetoma grain.}
\label{grains_cases}
\end{figure*}

\noindent The performance of any AI model depends on the precision of our manual segmentation. Therefore, this process is carried out meticulously to outline the grain borders accurately, avoiding the inclusion of other tissue components or inflammatory cells unless they are part of the grain itself. Additionally, the annotated grains must encompass the entire grain area, even if there were background or visible fractures within the grain. Figure \ref{grains_cases} displays sample microscopic images along with their annotated grains.

\section{MyData: Mycetoma Histology Microscopic Images Database}
The database contains a total of 864 microscopic tissue images from 142 patients. It is composed of 471 eumycetoma and 393 actinomycetoma images. The images are in JPEG format, while the ground truth annotations are in TIFF format. Table \ref{dist_img} gives the summary of database images.  

\begin{table}[h] 
	\centering
	\caption{Summary of images in MyData considering the mycetoma organisms.}
	\label{dist_img} 
	\resizebox{\columnwidth}{!}{%
		\begin{tabular}{|c|c|c|c|}
			\hline
			\textbf{Causative Agent} & \textbf{No. Images} & \textbf{Species} & \textbf{No. Images} \\ \hline \hline
			\multirow{5}{*}{Eumycetoma} & \multirow{5}{*}{471} & Madurella spp. (Mspp) & 149 \\ 
			& & MM+ & 167 \\ 
			& & MM- & 110 \\ 
			& & Aspergillus spp. (Aspp) & 36 \\ 
			& & Fusarium spp. (Fspp) & 9 \\ \hline
			\multirow{3}{*}{Actinomycetoma} & \multirow{3}{*}{393} & Actinomadura pelletieril (AMP) & 57 \\  
			& & Actinomadura madurae (AMM) & 62 \\ 
			& & Streptomyces somaliensis (SS) & 274 \\ \hline
		\end{tabular}
	}
\end{table}

\subsection{Images conditions}
Technical and human errors: \\
\noindent  The tissue samples were collected, labelled, prepared, and diagnosed by laboratory technicians and pathologists. Following collection, samples were processed and examined using histopathological techniques and grain culture, with results compared to verify accuracy. Any discrepancies led to the removal of samples from the database, and labels and results were double-checked to minimise errors.

\noindent  A challenge in mycetoma histopathological diagnosis is the complete or partial disappearance of grains in tissue slides, even though they are present in the tissue blocks. The sectioning or mounting process often causes this issue. The solid or rigid texture of mycetoma grains can lead to improper sectioning due to the thinness of the slides. To address this, we prepared two slides for each patient sample to maximize the chance of grain presence.

\noindent Folding and dissociation of sections in slides is a common problem in classical histology processing. Therefore, we accepted this as a normal condition in mycetoma tissue processing. Slides with folds were treated as regular slides and included in the database. From the two prepared slides, we selected the one that contained the grain and had a better appearance in terms of folding and tissue separation.

\subsection{Exclusion and inclusion criteria of samples in MyData}
\begin{figure*} 
	\centering
	\begin{subfigure}[b]{0.32\textwidth}
		\centering
		\includegraphics[height=0.18\textheight]{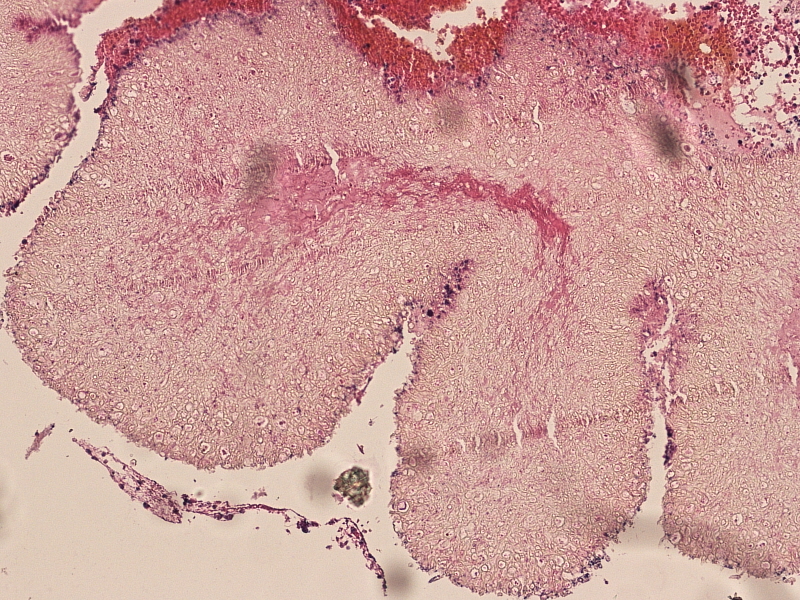}
		\caption{}
		\label{grains_exclude_1}
	\end{subfigure}%
	\hspace{0.01\textwidth} 
	\begin{subfigure}[b]{0.32\textwidth}
		\centering
		\includegraphics[height=0.18\textheight]{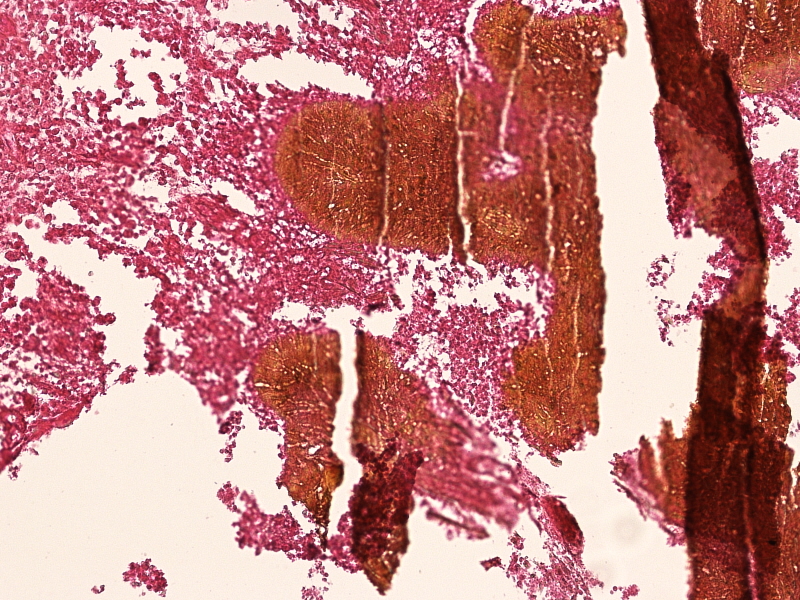}
		\caption{}
		\label{grains_exclude_2}
	\end{subfigure}%
	\hspace{0.01\textwidth} 
	\begin{subfigure}[b]{0.32\textwidth}
		\centering
		\includegraphics[height=0.18\textheight]{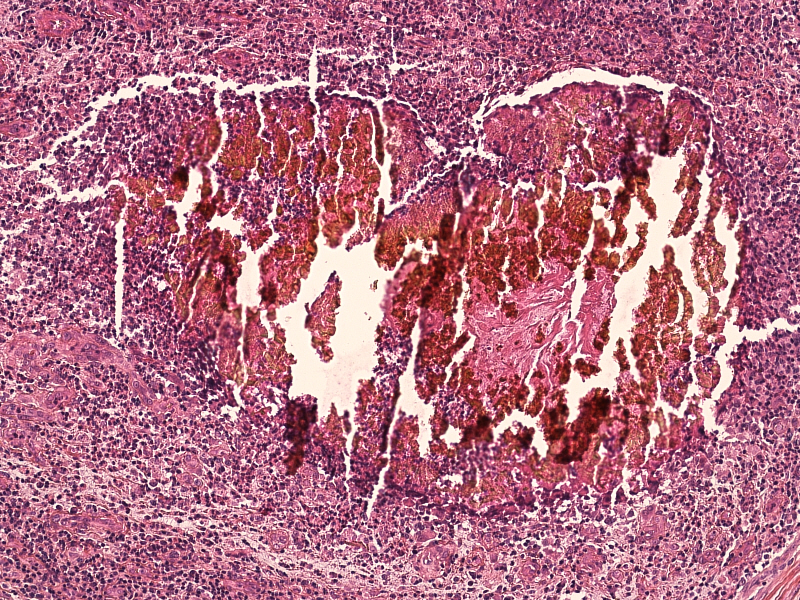}
		\caption{}
		\label{grains_exclude_3}
	\end{subfigure}
	\caption{Example of the images included in the database}
	\label{grains_exclude}
\end{figure*}

Considering the conditions mentioned, the following criteria were defined for including and excluding images:
\begin{itemize}
	\item For each patient, all grains in the tissue slide are included regardless of size and shape.
	\item If a grain is larger than the image size but fits the field, the visible part is included (Figure \ref{grains_exclude_1}).
	\item Partially folded grains (Figure  \ref{grains_exclude_2}) are included; completely folded grains are excluded.
	\item Tissue sections showing dissociation due to tissue reaction or processing (Figure  \ref{grains_exclude_3}) are included.
\end{itemize}

\subsection{Labelling}
Labels were defined considering patient ID to avoid statistical bias. Eumycetoma (fungus) and actinomycetoma (bacteria) are denoted as FM and BM, respectively. The naming format is main-type\_patient-ID\_grain-number. Ground-truth segmentation for grains adds a “\_mask” suffix to the grain image name. For example, \textit{FM3\_4} means it is an eumycetoma sample from patient 3, grain 4. The segmentation for FM3\_4 is named \textit{FM3\_4\_mask}.

\section{Conclusion}

This report describes the creation of MyData, the first database of mycetoma histopathological microscopic images. It includes detailed information on the samples used to compile the database. MyData is a comprehensive collection containing 864 microscopic tissue images with segmented ground-truth grain images from 142 patients. The database includes 80 eumycetoma and 62 actinomycetoma samples, representing four of the five most common mycetoma species worldwide. This makes MyData an important initial step toward developing a broader and larger database with various species from around the world, especially given the specific preparation and acquisition protocols provided.

\noindent MyData paves the way for global collaboration by encouraging laboratories worldwide to collect new samples. It also supports the creation of databases for different types of mycetoma images, such as X-rays and ultrasounds. We are open to collaboration and the exchange of knowledge, skills, and expertise to further advance this field. Additionally, we provided guidelines and optimised protocols for image preparation and acquisition. Following these guidelines offers numerous benefits, such as maintaining consistency, improving quality assurance, reducing errors, enhancing productivity, and preserving the integrity of the data.

\noindent Initially, the MyData database was collected and prepared to develop a diagnostic model for mycetoma. Recognising its potential value to the scientific community, we later decided to publish the database. The collected samples reflect the epidemiological distribution of mycetoma in Sudan, the epicenter of this disease. Given that mycetoma is a neglected disease, we believe the MyData database is highly suitable for publication. The database size is sufficient and comparable to the total infected population, and it includes the most common causative organisms in Europe and Africa, as well as less common ones in South America and Asia.

\noindent Using MyData, we developed and evaluated a machine learning model to analyze histopathological microscopic images of mycetoma grains and classify the disease as either eumycetoma or actinomycetoma. The model, which employs radiomics and partial least squares, achieved an accuracy of 91.89\% \citep{omar2024evaluation}. This illustrates the potential applications of the dataset in advancing diagnostic tools for mycetoma. 

\noindent  As the first database of mycetoma histopathological microscopic images, MyData represents a significant advancement. It provides a solid foundation for future research and can be expanded and refined in subsequent versions to further enhance its utility and comprehensiveness. In future work, we plan to assess inter- and intra-rater variability to strengthen the annotations further. This assessment will help evaluate the consistency and reliability of the annotations, ensuring the database's robustness for ongoing and future research.


\acks{The Ministry of Higher Education and Scientific Research, Sudan and Campus France supported this work. This study was also supported by the L'Oréal UNESCO For Women in Science and the European Mathematical Society (EMS).  We extend our gratitude to the MRC team for their assistance in data collection and annotation. A special acknowledgement goes to Mrs. Sahar Alhesseen, whose dedication and effort were crucial in managing and completing the AFRICAI repository data upload.  }

%
\ethics{This study received approval from the Soba University Hospital Institutional Review Board Committee in Khartoum, Sudan (No. SUH 05/01/2019). Clinical data were collected from patients treated at the Mycetoma Research Center (MRC). Prior to collecting surgical biopsies, written informed consent was obtained from each patient. These biopsies were taken from patients who visited the MRC over the past five years, representing a range of mycetoma types, durations, and clinical presentations.\\ }


\coi{We declare we do not have conflicts of interest. }

\data{The dataset is available in the AFRICAI repository \citep{MyData_AfricAI_2024} and Zenodo \citep{MyData_Zenodo_2024} under the terms of the Creative Commons Attribution (CC BY) license. Access to the dataset is restricted and requires submission of a formal request. The data adhere to the FAIR data principles and follow the guidance outlined in AFRICAI Imaging Repository \citep{miccai_africai_whitepaper_2024} for proper data uploading and backup procedures.}

\bibliography{sample}


\end{document}